\documentclass[aps,prc,amsfonts,preprintnumbers,superscriptaddress,showpacs,nofootinbib]{revtex4}

\usepackage{graphics}
\usepackage{psfrag}
\usepackage{epsfig}
\usepackage{psfig}
\usepackage{epsf}
\usepackage{float}

\newcommand{\fet}[1]{\mbox{\boldmath $#1$}}

\newcommand{\beq}{\begin{equation}}
\newcommand{\eeq}{\end{equation}}
\newcommand{\beqa}{\begin{eqnarray}}
\newcommand{\eeqa}{\end{eqnarray}}
\newcommand{\nn}{\nonumber \\ }

\newcommand{\no}{\nonumber}

\begin{document}



\title{On the renormalization of the one--pion exchange potential and 
the consistency of Weinberg's power counting\footnote{Dedicated to Professor Henryk Witala at the
occasion of his 60th birthday.}}


\author{E. Epelbaum}
\email[]{Email: evgeny.epelbaum@rub.de}
\affiliation{Institut f\" ur Theoretische Physik II, Fakult\" at f\" ur Physik und Astronomie,\\ Ruhr-Universit\" at Bochum 44780 Bochum, Germany}
\author{Ulf-G. Mei{\ss}ner}
\email[]{Email: meissner@hiskp.uni-bonn.de}
\homepage[]{URL: www.itkp.uni-bonn.de/~meissner/}
\affiliation{Universit\"at Bonn, Helmholtz-Institut f{\"u}r
  Strahlen- und Kernphysik (Theorie) and Bethe Center for Theoretical Physics, D-53115 Bonn, Germany}
\affiliation{Forschungszentrum J\"ulich, Institut f\"ur Kernphysik, 
Institute for Advanced Simulation and J\"ulich Center for Hadron Physics, D-52425 J\"ulich, Germany}


\date{\today}

\begin{abstract}
Nogga, Timmermans and van Kolck recently argued that Weinberg's power counting in the few--nucleon sector 
is inconsistent and requires modifications. Their argument is based on the observed cutoff dependence 
of the nucleon--nucleon scattering amplitude calculated by solving the Lippmann-Schwinger equation with 
the regularized one--pion exchange potential and the cutoff $\Lambda$ varied in the range $\Lambda = 2 \ldots 20$ 
fm$^{-1}$. In this paper we discuss the role the cutoff plays in the 
application of chiral effective field theory to the two--nucleon system and study carefully the cutoff--dependence of 
phase shifts and observables based on the one--pion exchange potential. We show that
(i) there is no need to use the momentum--space cutoff larger than $\Lambda
\sim 3$ fm$^{-1}$; 
(ii) the neutron--proton low--energy data show no evidence for an
inconsistency of Weinberg's power counting if one uses $\Lambda \sim
3$ fm$^{-1}$.
\end{abstract}


\maketitle


\vspace{-0.2cm}

\section{Introduction}
\def\theequation{\arabic{section}.\arabic{equation}}
\label{sec:intro}

The use of the chiral effective Lagrangian in the analysis of the nuclear forces
was pioneered by Weinberg~\cite{Weinberg:1990rz,Weinberg:1991um}. He realized
that a perturbative treatment like in the pion or the pion-nucleon sector is not 
possible -- the small nuclear binding energies require a non-perturbative
resummation. Weinberg thus proposed to apply the power counting -- the
necessary ingredient to organize any effective field theory -- to the
effective nucleon-nucleon (NN) potential and generate the nuclear bound and scattering
states utilizing this potential in a suitably regularized Lippmann-Schwinger
equation. This approach has enjoyed a considerable success in the description
of two-, three- and four-nucleon systems, for recent reviews see
Ref.~\cite{Bedaque:2002mn,Epelbaum:2005pn}. However, soon after Weinberg's seminal work, some
criticism has been raised about the consistency of this approach based  
on the role of the leading contact interaction $\propto M_\pi^2$
\cite{Kaplan:1998we,Beane:2001bc}. On the other hand, the
conclusions made in these papers have been questioned in Ref.~\cite{Gegelia:2004pz}.   
More recently, another issue has been raised in Ref.~\cite{Nogga:2005hy} 
(that paper also contains a detailed
discussion of earlier work on the consistency of Weinberg's approach 
and the corresponding references), see also \cite{Birse:2005um} for a related discussion.
In particular, it was argued that  cutoff independence for a larger range
of cutoff values than usually employed requires the promotion of certain
next--to--leading order (NLO)
operators to the leading order (LO). At this order, the effective potential
is given by the one--pion exchange and two S--wave four--nucleon contact
terms in Weinberg's power counting. In the modified counting of
Ref.~\cite{Nogga:2005hy}, three additional contact interactions in P-- and
D--waves are included in the LO potential. Here, we critically assess the statements
made  in that paper and draw some different conclusions, simply because we
provide a {\em quantitative} measure of the theoretical accuracy following the
seminal work of Lepage~\cite{Lepage:1997,Lepage:2000}. In particular, the
low--energy neutron--proton scattering data show no evidence for an
inconsistency of Weinberg's original scheme.
At this point, we
should already stress that such investigations  based on the LO potential are limited
in their significance --- the two-pion exchange potential that only enters at NLO is
a very important part of the nuclear force. The well-known and important cancellations
between pion- and $\rho$-exchange in the tensor channel could be interpreted
as a signal that it also should be treated nonperturbatively. Even though effective
field theory is the appropriate method to deal with the problem of nuclear
forces, much has already been learned in less systematic schemes (like
e.g. meson exchange models) and this knowledge needs to be accounted for.

 Our manuscript is organized as follows. Section~\ref{sec:regul} contains a
thorough discussion of the issues underlying regularization and renormalization 
of the Lippmann--Schwinger equation. In Sec.~\ref{sec:phases}, we analyze
the cutoff dependence of the NN phase shifts, followed by a detailed
comparison of observables calculated in the original Weinberg scheme and
the modified power counting proposed in Ref.~\cite{Nogga:2005hy} in 
Sec.~\ref{sec:observ}. A generalization of these issues  to higher orders
is discussed in Sec.~\ref{sec:discussion}. We conclude and summarize our
results in Sec.~\ref{sec:summary}.

\section{Regularization and renormalization of the Lippmann--Schwinger equation}
\def\theequation{\arabic{section}.\arabic{equation}}
\label{sec:regul}

The leading--order two--nucleon force is given by the one--pion exchange potential (OPEP) accompanied 
by contact interactions (counterterms). The choice of counterterms is what makes the difference 
between the original \cite{Weinberg:1990rz,Weinberg:1991um} and modified \cite{Nogga:2005hy} Weinberg power counting schemes. 
The well--known expression for the LO one--pion exchange potential in momentum space reads 
\beq
\label{opep_mom}
 V_{1\pi} ( \vec q \, )  
= -\biggl(\frac{g_A}{2F_\pi}\biggr)^2 \,
\fet \tau_1 \cdot \fet \tau_2 \,
\frac{\vec{\sigma}_1 \cdot\vec{q}\,\vec{\sigma}_2\cdot\vec{q}}
{\vec q \, ^2 + M_\pi^2}\,,
\eeq
where $\vec q = \vec p \, ' - \vec p$ is the nucleon momentum transfer and $\vec \sigma_i$ ($\fet \tau_i$) 
are spin (isospin) matrices of the nucleon $i$. Further, $g_A$, $F_\pi$ and $M_\pi$ denote the 
axial--vector coupling constant, pion decay constant and pion mass, respectively.
In this work, we use the values $F_\pi = 92.4$ MeV 
and $M_\pi = 138.03$ MeV. For the axial coupling constant $g_A$, we take the same value $g_A=1.29$ 
as in our previous studies \cite{Epelbaum:2003xx,Epelbaum:2004fk} which accounts for the 
Goldberger--Treiman discrepancy and is, therefore, larger than the empirical value of $g_A=1.267$. 
Both values for this LEC can be used at LO since the difference between them is a higher--order effect.  
In coordinate space, the OPEP is local and takes the form
\beq
\label{opep_coord}
 V_{1\pi} ( \vec r \, )  
= \biggl(\frac{g_A}{2F_\pi}\biggr)^2 \,
\fet \tau_1 \cdot \fet \tau_2 \,
\left[M_\pi^2\, \frac{e^{- M_\pi r}}{12 \pi r}  \,\left( S_{12} (\hat r ) \, \left( 1 + \frac{3}{M_\pi r} + \frac{3}{(M_\pi r )^2} \right)
+  \vec \sigma_1 \cdot \vec \sigma_2 \right) - \frac{1}{3} \, \vec \sigma_1 \cdot \vec \sigma_2 
\, \delta^3 (r )
\right]\,,
\eeq
with $\vec r$ being the relative distance between the nucleons, $r = | \vec r \, |$, $\hat r = \vec r/r$ and 
\beq
S_{12} = 3 \, \vec \sigma_1 \cdot \hat r \, \vec \sigma_2 \cdot \hat r - \vec \sigma_1 \cdot \vec \sigma_2\,.
\eeq
Notice that at higher orders in chiral EFT, the OPEP receives isospin--conserving 
and isospin--violating corrections, see e.g.~\cite{Epelbaum:2005fd}, as well as the corrections 
of relativistic origin \cite{Friar:1999sj}.

The nonrelativistic Lippmann--Schwinger (LS) equation for the half--off--energy shell $T$--matrix is given by 
\beq
\label{LS}
T (\vec p \, ', \, \vec p \, ) = V (\vec p \, ', \, \vec p \, )  + \int d^3 p '' \; 
V (\vec p \, ', \, \vec p \, '' ) \frac{m}{\vec p \, ^2 -   { \vec p \, '' }^2 + i \epsilon} 
T (\vec p \, '', \, \vec p \, )\,,
\eeq
with $m$ being the nucleon mass, $\vec p$ the on--shell point and $V$ the potential. 
Clearly, the OPEP in Eq.~(\ref{opep_mom}) as well as contact interactions lead to ultraviolet 
divergencies in the LS equation which, therefore, needs to be regularized. This is usually achieved 
by multiplying the potential  $V (\vec p, \; \vec p \, ')$  with a regulator function 
$f^\Lambda$,
\beq
\label{pot_reg}
V (\vec p, \; \vec p \, ') \rightarrow f^\Lambda ( \vec p \, ) \, 
V (\vec p, \; \vec p \, ')\, f^\Lambda ( \vec p \,' )\,. 
\eeq 
An exponential regulator function 
\beq
\label{reg_fun}
f^\Lambda ( \vec p \, ) = \exp [- \vec p \,^{2n}/\Lambda^{2n} ]
\eeq 
with $n \geq 1$ is the commonly used choice. In particular, $f^\Lambda ( \vec p \, )$  
with $n=2$ was employed in Refs.~\cite{Epelbaum:1999dj,Epelbaum:2002ji,Nogga:2005hy} while $n=3$ was used in 
\cite{Epelbaum:2003xx,Epelbaum:2004fk}.

Before discussing the power counting, let us first take a closer look at the LS equation 
for the pure OPEP in the limit $\Lambda \to \infty$. It is easy to see that no ultraviolet 
divergences show up in the spin--singlet channels (with the exception of the
$^1S_0$ partial wave), where the OPEP takes the form
\beq
\label{opep_singl}
V_{1\pi}^{S=0} ( \vec q \, )  
= \biggl(\frac{g_A}{2F_\pi}\biggr)^2 \,
\fet \tau_1 \cdot \fet \tau_2 \biggl( 1 -
\frac{M_\pi^2} {\vec q \, ^2 + M_\pi^2} \biggr)\,.
\eeq
Despite the fact that the individual terms in the series resulting from iterating the LS equation (\ref{LS}) 
are ultraviolet divergent in all spin--triplet channels, the equation can still be solved nonperturbatively 
for those uncoupled partial waves in which the singular part of the OPEP is repulsive. 
The resulting phase shifts are uniquely determined. The easiest way to see this is by looking at the radial 
Schr\"odinger equation which possesses a unique regular solution for repulsive singular potentials. 
On the contrary, it is ill--defined in the 
uncoupled channels where the $1/r^3$--part of the OPEP is
attractive.
For these partial waves, both solutions 
of the radial Schr\"odinger equation go to zero at the origin while oscillating infinitely rapidly  \cite{Case:1950an}. 
Imposing orthogonality of solutions  allows to 
fix the discrete spectrum (which is unbound from below) and scattering states in terms of a single 
energy--independent phase factor \cite{Case:1950an}, see also \cite{Frank:1971aa} for a comprehensive 
review on singular potentials. The case of coupled channels with $l, l' = j \pm 1$ can be reduced to the 
uncoupled one by decoupling the two Schr\"odinger equations in the
short--distance region 
\cite{Sprung:1994aa,PavonValderrama:2005gu,Beane:2001bc}. 
For the OPEP, this yields the decoupled equations with an attractive and repulsive singular potentials for all $j$, 
so that one always needs to specify a single constant in order to uniquely determine both phase shifts and the mixing angle.    

In practice, the LS equation (\ref{LS}) is usually studied for finite values of the cutoff $\Lambda$. 
Although the scattering amplitude based on the regularized OPEP is well--defined for any arbitrarily large but finite 
$\Lambda$, the limit $\Lambda \to \infty$ only exists in spin--singlet and repulsive uncoupled spin--triplet channels.  
In all coupled and uncoupled attractive spin--triplet channels the amplitude shows oscillatory behavior when 
$\Lambda$ is taken to infinity. It is known that the amplitude may be stabilized in each of these channels 
(for a finite energy) by introducing a single local counterterm\footnote{The term ``local'' means in this context that 
the corresponding nonregularized operator acts only near the origin.} and adjusting its strength to reproduce 
e.g.~the corresponding scattering length
\cite{Beane:2000wh,Beane:2001bc,Nogga:2005hy}.\footnote{This was demonstrated
  analytically in coordinate space for particular choices of
  regularizations, see e.g.~\cite{PavonValderrama:2005gu}, and confirmed
  numerically in momentum-space calculations in
  Refs.~\cite{Meissner:2001aa,Nogga:2005hy} using different
  regularizations.  This suggests the universality, i.e.~independence on a
  particular regularization prescription, of the above observation.} 
That is, a well--defined limit $\Lambda \to \infty$ 
of the scattering amplitude exists provided  the strength $C (\Lambda )$ of the counterterm is 
fixed as described above. A similar observation was also made for the chiral potential at next--to--next--to--leading order 
in selected channels \cite{Meissner:2001aa}. For a given partial wave, the above procedure provides the way
to explicitly remove the scale associated with the ultraviolet cutoff and appears to be similar to the 
standard renormalization procedure applied in (perturbative) quantum field theory calculations. 
Despite this formal similarity, an important difference is given by the fact that the 
quantum mechanical approach described above is intrinsically nonperturbative. 
S--matrix elements depend nonanalytically on the strength of the OPEP for $\Lambda \to \infty$
in all spin--triplet channels  \cite{Frank:1971aa} which negates a
perturbative description in the weak--coupling limit.\footnote{A similar
  observation is made in Ref.~\cite{Beane:2000wh} for the case of S-waves.}. 

We now turn to the actual subject of this study and consider the two--nucleon system at LO in chiral EFT 
based on Eq.~(\ref{LS}) with the potential given by Eq.~(\ref{opep_mom}) and accompanied by local counterterms. 
Let us first ignore the potential interpretation difficulty due to the lack of the weak--coupling limit 
and require the scattering amplitude to be explicitly independent on the ultraviolet cutoff $\Lambda$.
That is, we take the limit $\Lambda \to \infty$. As follows from the above discussion, an {\it infinite} number 
of local counterterms (more precisely, one term in each coupled and uncoupled attractive spin--triplet channels)  
are then needed in order to make the scattering amplitude well--defined. While such an approach still has
predictive power for individual partial waves, the fact that an infinite number of parameters need to be 
determined to describe e.g.~the cross section for a given energy makes it useless for practical applications.    
It has been argued in Ref.~\cite{Nogga:2005hy} that this complication might be avoided if the OPEP is treated 
in perturbation theory for high partial waves. We do not consider this argument as being relevant for the 
present discussion for the following reasons. First of all, it is disturbing that one needs to rely on 
a particular computational technique, namely the partial wave decomposition, in order to solve  Eq.~(\ref{LS}).
This would rule out e.g.~the three--dimensional approach of Ref.~\cite{Fachruddin:2000wv}.  
Secondly, if one decides to treat all partial waves for $l, l' \geq l_{\rm crit}$ in perturbation theory,
that is, if one assumes that $l_{\rm crit}$ is a large number, $l_{\rm crit} \gg 1$, it is hard to argue 
why partial waves with $l, l' = l_{\rm crit} -1$ should not be treated
perturbatively as well, 
see, however, Ref.~\cite{Birse:2005um} for a recent discussion based on
secular perturbation theory. Most important,
however, is the fact that the applicability of perturbation theory for high partial waves relies on 
the repulsive effect of the centrifugal barrier that goes like $l (l+1)/r^2$ and is only justified for regular 
potentials of a finite range. This certainly does not apply to attractive singular $1/r^3$--potentials.
The perturbative treatment of high partial waves is formally only justified if a finite cutoff is used in the LS equation.
This possibility will be discussed below. In the limit of an infinitely large cutoff $\Lambda$, no predictions can be made 
for scattering observables (such as e.g.~the cross section) if only a finite number of experimental data is 
available (i.e. if one does not know a priori that phase shifts are small for {\it all} high partial waves).  

Clearly, the need to include infinitely many counterterms to absorb the ultraviolet 
divergences in the LS equation based on the OPEP is a direct consequence of the nonperturbative 
nature of the problem at hand. Contrary to standard chiral perturbation theory in the Goldstone Boson 
and single--nucleon sectors, where diagrams with a fixed number of loops contribute at a given order 
in the chiral expansion, infinite number of loops need to be taken into account at LO in the NN case. 
Taking the chiral limit of the OPEP, it is easy to verify based on the dimensional analysis 
that counterterms of the order $Q^{2n}$ with $n=1,2,\ldots$ and $Q$ being the generic low--momentum scale 
are required to absorb the divergences arising from $2n$ iterations of the LS equation, see 
\cite{PavonValderrama:2005uj} for a similar observation made based on the coordinate--space representation.  
While this difficulty might first appear as an indication of the uselessness of the 
chiral EFT approach in the few--nucleon sector, it is actually not. 
Contrary to renormalizable field theories like QCD, 
EFTs (such as e.g.~pionless or chiral EFT) are low--energy expansions with the finite radius of 
convergence representing their intrinsic ultraviolet cutoff  $\bar \Lambda$.
As a consequence, no improvement in the description of the data can, 
generally,  be expected increasing $\Lambda$ beyond the pertinent hard scale $\bar \Lambda$. 
Keeping $\Lambda$ finite does, therefore, not mean any loss of generality 
(provided the error due to $\Lambda$ being finite is within the theoretical uncertainty at a given order)
neither does it violate the requirement on the EFT of being a systematically 
improvable low--energy approximation to a given underlying theory. 

The above arguments make it clear that nonrenormalizability of the OPEP in Eq.~(\ref{opep_mom}) is 
not a fundamental problem but rather an artifact of a particular choice for the 
large--momentum (or, equivalently, short--distance) behavior of the OPEP out of infinitely 
many {\it equivalent} (from the viewpoint of chiral EFT) ones. For example, including the 
corrections to account for the proper normalization of the nonrelativistic single--nucleon 
states yields the expression for the OPEP of the form
\beq
\label{opep_rel}
\frac{m}{E} \, V_{1\pi} (\vec q \, ) \,  \frac{m}{E} = \frac{m}{\sqrt{\vec p \, '\,^2 + m^2}} \, V_{1\pi} (\vec q \, ) \,  
\frac{m}{\sqrt{\vec p \, ^2 + m^2}} \,,
\eeq
which does not lead to ultraviolet divergences in the LS equation. Despite the different properties in the 
ultraviolet region, both forms in Eq.~(\ref{opep_mom})
and Eq.~(\ref{opep_rel}) are, of course, equally valid (as well as the expression resulting from the $p/m$--expansion 
of Eq.~(\ref{opep_rel})) as the differences between them are of higher orders in the EFT expansion. 
The behavior of the potential for momenta larger than $\bar \Lambda$    
is not constrained in chiral EFT and can always be modified (via inclusion of a set of higher--order terms) to 
make the LS equation free from ultraviolet divergences, see e.g.~Eq.~\ref{pot_reg}. 

Having accepted the viewpoint that applying the limit $\Lambda \to \infty$ does not represent a  
fundamental requirement in EFT and might even not be useful for
chiral EFT applications in the few--nucleon sector, 
the following two issues need to be clarified and will be dealt with in the next sections:
\begin{itemize}
\item
how to choose the value of $\Lambda$ to make the EFT expansion most efficient~?
\item
what counterterms need to be taken into account, i.e.~what are the implications for the power counting~?
\end{itemize}

\section{Cutoff dependence of the NN phase shifts}
\def\theequation{\arabic{section}.\arabic{equation}}
\label{sec:phases}

In the past decade, the NN system has been explored within the chiral EFT 
framework in Weinberg's formulation \cite{Weinberg:1990rz,Weinberg:1991um} based on the finite momentum--space cutoff 
$\Lambda$ in the range $300 \ldots 700$ MeV in \cite{Park:1998cu},
$450 \ldots 650$ $(600)$ MeV in \cite{Epelbaum:2003xx} (\cite{Epelbaum:2004fk}), $500$ MeV in \cite{Entem:2003cs}, 
$500 \ldots 1000$ MeV in \cite{Ordonez:1995rz,Epelbaum:1999dj}. In addition, much larger cutoff values upto $\sim 4$ GeV 
were used in the LO calculation of Ref.~\cite{Nogga:2005hy}, where certain contact terms of higher--orders in the 
original Weinberg power counting scheme have been included. Finally, the case of an infinite cutoff  
has been studied in Refs.~\cite{PavonValderrama:2004nb,PavonValderrama:2005gu,Valderrama:2005wv,PavonValderrama:2005uj}
using the boundary condition regularization prescription. 
We further emphasize that chiral two-pion exchange has also been tested in the
the Nijmegen partial wave analysis using the coordinate-space cutoffs $R=1.4$, $1.6$
and $1.8$ fm \cite{Rentmeester:1999vw,Rentmeester:2003mf}. 
This large variation in values for $\Lambda$ adopted 
by different authors raises the important question on the criteria for choosing the cutoff. Clearly, taking 
$\Lambda$ too small, i.e.~much smaller than the separation scale $\bar \Lambda$, will remove the truly 
long--distance physics and reduce the predictive power of the theory. On the other hand, it is argued in 
\cite{Lepage:1997,Lepage:2000} that too large values for $\Lambda$ result in a highly nonlinear 
behavior and should be avoided as well, see also \cite{Beane:2000fx} and
\cite{Gegelia:2004pz}. 
In order to clarify this last statement, 
let us make a rather plausible assumption that the low--energy two--nucleon
dynamics 
can be described within the 
framework of quantum mechanics based on the Hamilton operator given by the kinetic energy and a non--singular NN 
potential (in practice, one can think about e.g.~the Nijm I, II or Reid 93 
potentials \cite{Stoks:1994wp}). Keeping the point--like static OPEP 
in Eq.~(\ref{opep_mom}) as the leading approximation of the long--range part of the underlying interaction,
it is certainly preferable to adopt the smallest (yet acceptable) value for the cutoff,
that is, $\Lambda \sim \bar \Lambda$. Increasing $\Lambda$ substantially beyond this scale would result in 
incorporating the incorrect singular short--range part of the OPEP and, as a
consequence, would require 
fine-tuning of a large number of counterterms to compensate for
  it.\footnote{The precise number of parameters that need to
  be fine tuned depends on the cutoff choosen.} 
In the extreme limit $\Lambda \to \infty$, fine-tuning of an infinite 
number of counterterms is necessary. 

Based on the above arguments, one might expect the ultraviolet cutoff of the order of the chiral 
symmetry breaking scale, i.e. $\Lambda_\chi \sim 1$ GeV, to be a reasonable choice. On the other hand, 
the radius of convergence of the chiral expansion for the NN scattering amplitude cannot exceed  
$| \vec p \, | \sim M_\rho /2 \sim 400$  MeV since e.g.~the left--hand cut associated with the $\rho$--meson exchange
cannot be properly reproduced in chiral EFT at any finite order. A
(numerically) similar estimation follows also from the fact that the
formulation of chiral EFT based on the nucleon--nucleon potential 
with no explicit pions is only applicable below the pion production
threshold. These arguments suggest that even significantly smaller 
values of $\Lambda$  in Eq.~(\ref{reg_fun}) might be sufficient. A detailed discussion on the 
choice of ultraviolet cutoff and its role in renormalization of the Schr\"odinger equation is 
given by Lepage \cite{Lepage:1997,Lepage:2000}. He argued that the coordinate--space (momentum--space) 
cutoff should not be decreased (increased) beyond the critical value, after which the description of 
the data stops to improve. Taking the cutoff near this critical point is the most efficient choice. 

\begin{figure}[tb]
  \begin{center} 
\includegraphics[width=16.6cm,keepaspectratio,angle=0,clip]{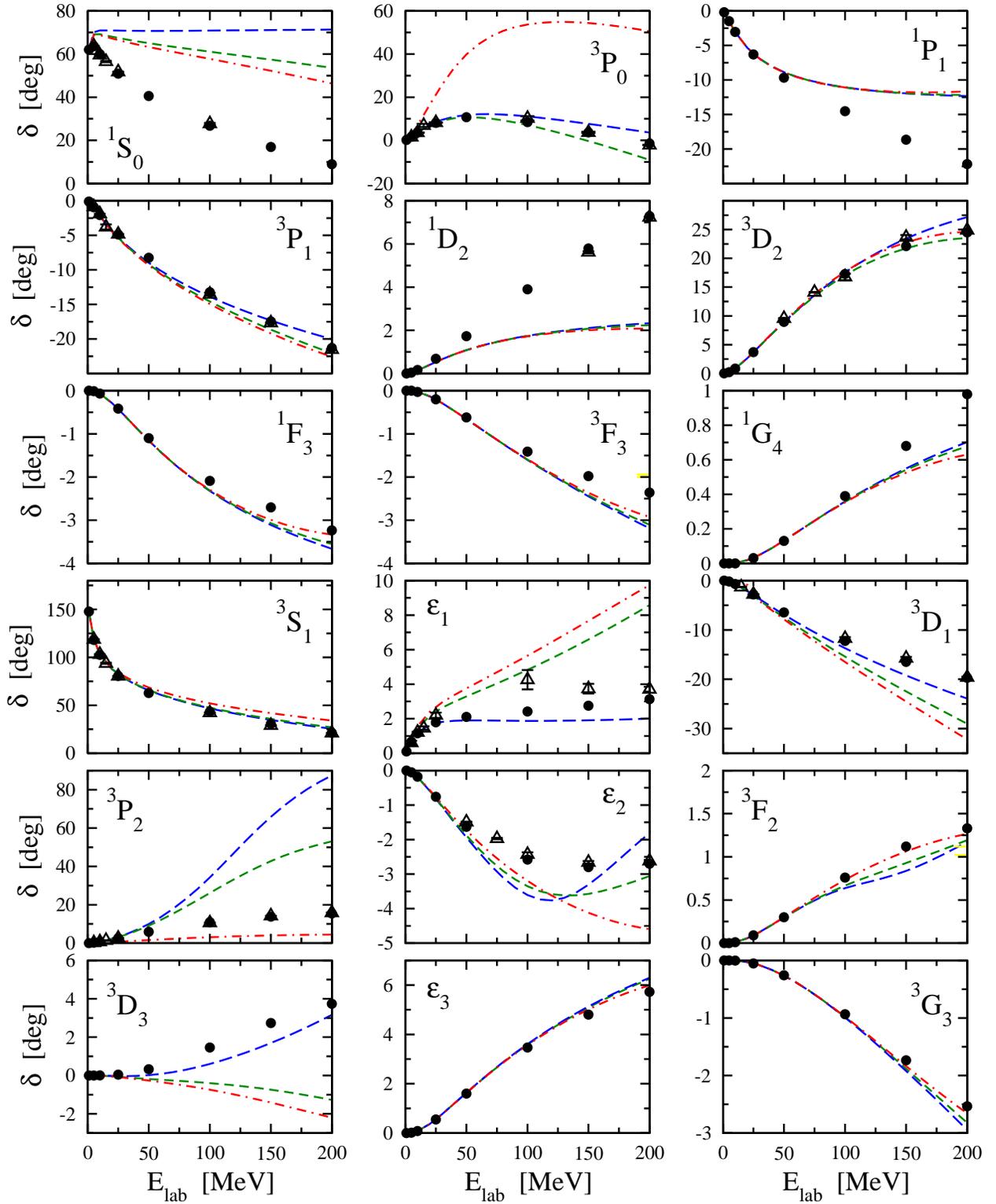}
    \caption{
         Comparison of the {\it np} phase shifts at LO in the original and modified Weinberg power counting 
         scheme to the Nijmegen \cite{Stoks:1993tb,NNonline} (solid dots) and Virginia Tech single--energy PWA \cite{SAID} (open 
         triangles). Dot--dashed lines correspond to the original Weinberg power counting with $\Lambda=2.5$ fm$^{-1}$
         while short--dashed (long--dashed) lines depict the results obtained within the modified counting scheme 
         with $\Lambda = 3$ fm$^{-1}$ ($\Lambda = 20$ fm$^{-1}$).
\label{phases} 
 }
  \end{center}
\end{figure}

In the present work, we determine {\em the minimal
acceptable value for the cutoff $\Lambda$ as the one for which  
the error due to keeping $\Lambda$ finite is within the theoretical uncertainty of the LO approximation.} 
To that aim, we use the modified power counting scheme introduced by Nogga et al.~\cite{Nogga:2005hy} which 
is compatible with the variation of $\Lambda$ in a large range we are interested in. We repeat the analysis of 
Ref.~\cite{Nogga:2005hy} and consider the same partial waves based on the OPEP in Eq.~(\ref{opep_mom})
accompanied by the counterterms and multiplied with the regulator function $f^\Lambda ( \vec p \, )$ in Eq.~(\ref{reg_fun}). 
We use the cutoff $\Lambda < 20$ fm$^{-1}$ 
$\sim 4$ GeV and include the counterterms following the rules of Ref.~\cite{Nogga:2005hy}, see also the 
discussion in section \ref{sec:regul}. In particular, one contact interaction is taken into account in the $^1S_0$ 
partial wave while no counterterms are included in the spin--singlet channels $^1P_1$, $^1D_2$, 
$^1F_3$ and $^1G_4$ as well as in the uncoupled spin--triplet partial waves $^3P_1$ and $^3F_3$ where 
the tensor part of the OPEP is repulsive. Further, one contact interaction is included in each of the coupled 
channels $^3S_1$--$^3D_1$ and $^3P_2$--$^3F_2$\footnote{In a coupled channel with the total angular 
momentum $j$, we take into account the leading counterterm in the partial wave with $l = l' = j-1$.} as well as in the 
uncoupled spin--triplet channels $^3P_0$ and $^3D_2$,
where the tensor part of the OPEP is attractive. Specifically, the following contact interactions are 
incorporated\footnote{The explicit separation of contact interactions in to
  $M_\pi$-independent and  $M_\pi$-dependent ones is irrelevant for the
  present investigation and, therefore, not considered.}:
\beqa\label{VC}
\langle S | V_{\rm cont}| S \rangle&=& C_{S}\,, \quad S=\{^1S_0, \; ^3S_1\} \,, \nn
\langle P | V_{\rm cont}| P \rangle&=& C_{P} \; p \, p'\,,\quad   P=\{^3P_0, \; ^3P_2\} \,,  \nn
\langle D | V_{\rm cont}| D \rangle&=& C_{D} \; {p}^2\,{p}'^2\,,\quad   D= {^3D}_2 \,.
\eeqa
For each value of the cutoff $\Lambda$ in Eq.~\ref{reg_fun}, the LECs $C_i$ are  
determined by fitting the corresponding phase shifts to the Nijmegen PWA \cite{Stoks:1993tb}
in the low--energy region $E_{\rm lab} < 10$ MeV. Our way of fixing the LECs is, therefore, 
somewhat different to the one adopted in \cite{Nogga:2005hy} and is more in spirit of 
Ref.~\cite{PavonValderrama:2004nb,Valderrama:2005wv,PavonValderrama:2005uj}. The results for 
the phase shifts and mixing angles are presented in Fig.~\ref{phases} for two choices of the 
cutoff: $\Lambda =$3 and 20 fm$^{-1}$.
We also show the results for the case of the $^3D_3$--$^3G_3$ coupled channels, where we do not include 
counterterms following Ref.~\cite{Nogga:2005hy}. 
The small differences between our results and the ones of Nogga et al.~may be attributed to the 
different fitting procedure, the choice of the regulator function and the
adopted value for $g_A$. 
The results for phase shifts with $\Lambda = 20$ fm$^{-1}$ are also very
similar to the LO results of Ref.~\cite{PavonValderrama:2005uj}. 
Notice that the obtained binding energies of deeply bound states in the
attractive spin--triplet channels are similar to the ones found in Ref.~\cite{Nogga:2005hy}. 
In particular, for $\Lambda = 20$ fm$^{-1}$ we find $E_{\rm b} =1.69$ and 34.27 GeV in the $^3S_1$--$^3D_1$,
0.22 and 21.48 GeV in the $^3P_0$ and 0.26 GeV in the $^3D_2$. While no deeply bound states 
appear in the $^3P_2$--$^3F_2$ and $^3D_3$--$^3G_3$ for $\Lambda \leq 20$ fm$^{-1}$, one should keep in mind that 
they necessarily show up for larger cutoff values. We further emphasize that
the binding energies of the first spurious states in the $^3P_0$ and $^3D_2$
channels are consistent with the above estimation for the separation scale
$\bar \Lambda$. 
Notice finally that the scattering amplitude might also
have spurious poles in the complex energy plane which are not associated with 
bound states.   

\begin{figure}[tb]
  \begin{center} 
\includegraphics[width=16.6cm,keepaspectratio,angle=0,clip]{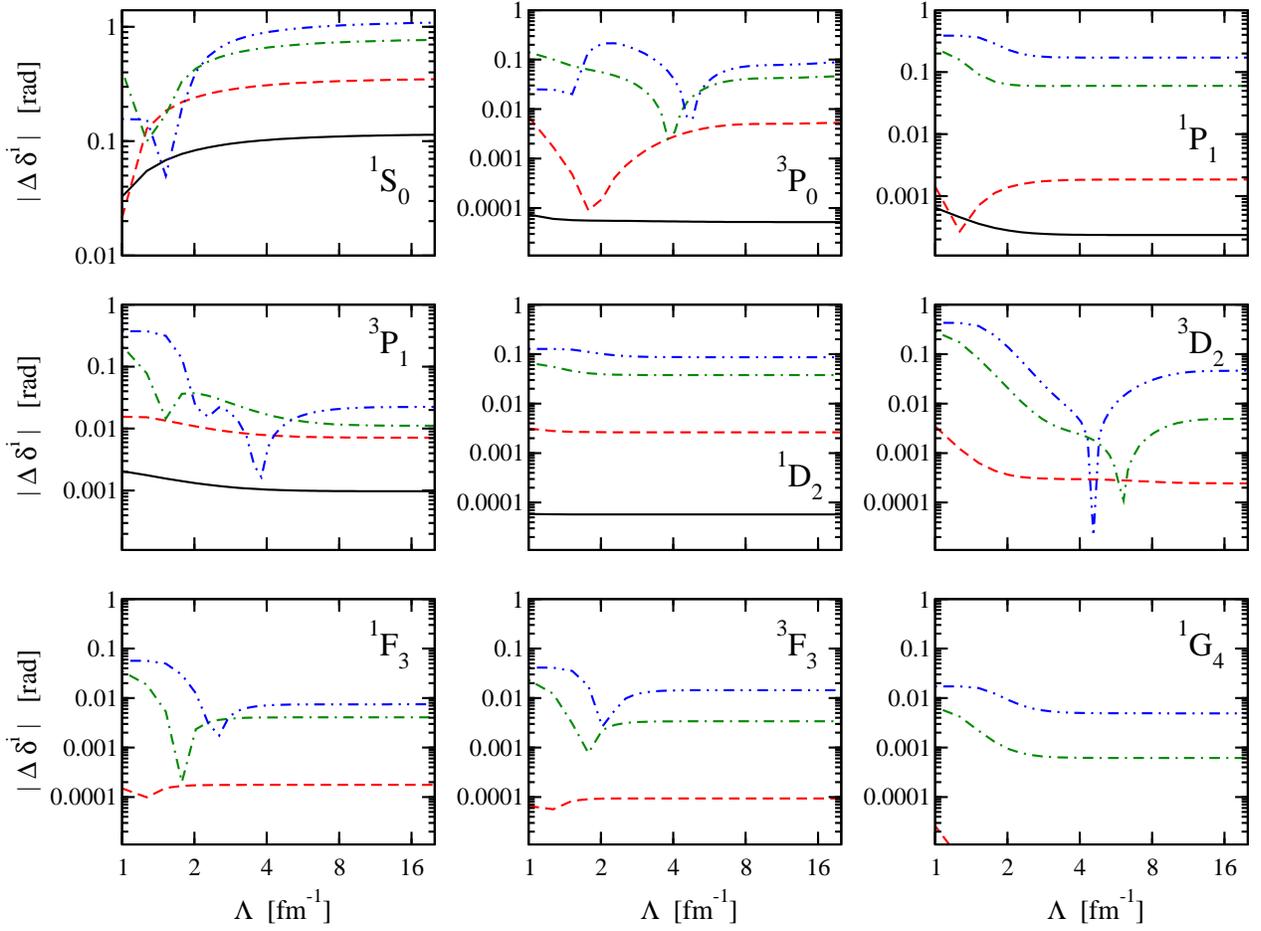}
    \caption{
         Error in phase shifts  in the uncoupled channels versus the cutoff $\Lambda$ 
         in the Lippmann--Schwinger equation. Results are given for laboratory energies of 5 MeV (solid line), 
         25 MeV (short--dashed line), 100 MeV (dot--dashed line) and 200 MeV (dot--dot--dashed line). 
\label{err_uncoup} 
 }
  \end{center}
\end{figure}

We are now in the position to determine the minimal acceptable value for $\Lambda$.  
To that aim, we very the cutoff in the range $\Lambda = 1 \ldots 20$ fm$^{-1}$ 
and study carefully the cutoff dependence of the phase shifts. 
Notice that since unstable channels are stabilized 
by adding the appropriate counterterms and tuning their strengths as described
above, all phase shifts and mixing angles are expected
to reach stable values as $\Lambda$ goes to infinity. 
For partial waves considered in this work, phase shift become nearly cutoff--independent
for $\Lambda > 20$ fm$^{-1}$. Thus, the results for $\Lambda = 20$ fm$^{-1}$
maybe regarded the ones in the limit $\Lambda \to \infty$. 
Clearly, as $\Lambda$ is taken very small, $\Lambda \sim 1$ 
fm$^{-1}$, one expects large errors in the description of the data. Increasing the cutoff values the 
results are expected to improve until $\Lambda$ reaches the critical point where missing higher--order contributions 
to the potential due to e.g.~$2\pi$--exchange become the dominant source of errors. Starting from this point, 
the error due to keeping $\Lambda$ finite is within the theoretical uncertainty of the LO approximation.
These qualitative expectations are confirmed by explicit calculations as shown
in Fig.~\ref{err_uncoup},
where errors in phase shifts $| \Delta \delta^i | = | \delta^i -
\delta^i_{\rm Nijm \; PWA} |$ in the uncoupled channels are depicted.  
Clearly the theoretical uncertainty at a given order in the EFT expansion cannot
be smaller than the deviation from the data. The lower bound for the LO
theoretical uncertainty in a partial wave $i$ is, therefore, given by $\lim_{\Lambda \to
  \infty} | \Delta \delta^i |$ and can be read off from Fig.~\ref{err_uncoup}. 
One observes that already for $\Lambda \gtrsim 3$ fm $^{-1}$ the error due to keeping
$\Lambda$ finite is within the theoretical uncertainty of the LO
approximation. The only exception is given by the $^3P_1$ phase shift, where 
the critical value of the cutoff $\Lambda \sim$ 4 -- 5 fm$^{-1}$ is somewhat
larger. On the other hand, 
even a smaller cutoff value $\Lambda \sim 2$ fm$^{-1}$ seems to be sufficient in the
$^1S_0$, $^1F_3$ and $^3F_3$ partial wave.  

\begin{figure}[tb]
  \begin{center} 
\includegraphics[width=6.0cm,keepaspectratio,angle=0,clip]{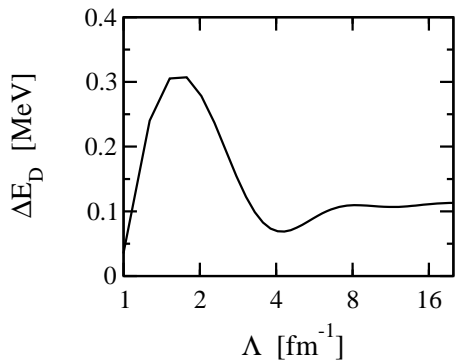}
    \caption{
         Error in the deuteron binding energy as a function of the cutoff $\Lambda$. 
\label{err_deut} 
 }
  \end{center}
\end{figure}

We now turn to the coupled channels. 
\begin{figure}[tb]
  \begin{center} 
\includegraphics[width=16.6cm,keepaspectratio,angle=0,clip]{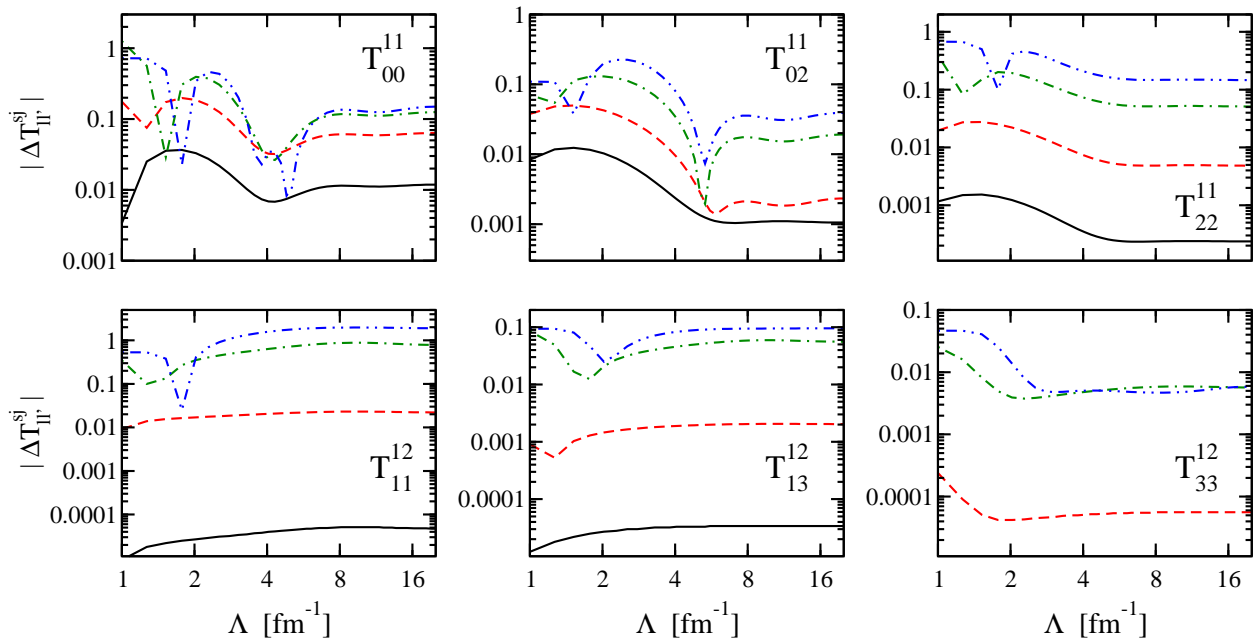}
    \caption{
         Error in T--matrix elements in the coupled channels versus the cutoff $\Lambda$ 
         in the Lippmann--Schwinger equation. Results are given for laboratory energies of 5 MeV (solid line), 
         25 MeV (short--dashed line), 100 MeV (dot--dashed line) and 200 MeV (dot--dot--dashed line). 
\label{err_coup} 
 }
  \end{center}
\end{figure}
Here we have decided to look at the error
in the T--matrix elements rather than in the corresponding phase shifts and mixing
angles in order to avoid the ambiguity associated with different
parametrizations of the coupled--channel S--matrix elements. More precisely,
we consider dimensionless on--the--energy--shell T--matrix elements
$T_{ll'}^{sj}$ related to S--matrix via 
\beq
T_{ll'}^{sj} = S_{ll'}^{sj} - \delta_{ll'} \,.
\eeq
Our results for the
$^3S_1$--$^3D_1$ and $^3P_2$--$^3F_2$ channels are shown in
Fig.~\ref{err_coup}. The critical values of the cutoff are $\Lambda \sim 2$ fm$^{-1}$
for $T_{11}^{12}$, $T_{13}^{12}$ and $T_{33}^{12}$,  $\Lambda \sim 3$ fm$^{-1}$ for $T^{11}_{00}$
and  $\Lambda \sim $ 4 -- 5 fm$^{-1}$ for $T^{11}_{02}$ and $T^{11}_{22}$. We
have also looked at the deuteron binding energy, see Fig.~\ref{err_deut},
which shows a similar pattern with the critical cutoff value $\Lambda \sim 3$
fm$^{-1}$. 

\begin{figure}[tb]
  \begin{center} 
\includegraphics[width=12.0cm,keepaspectratio,angle=0,clip]{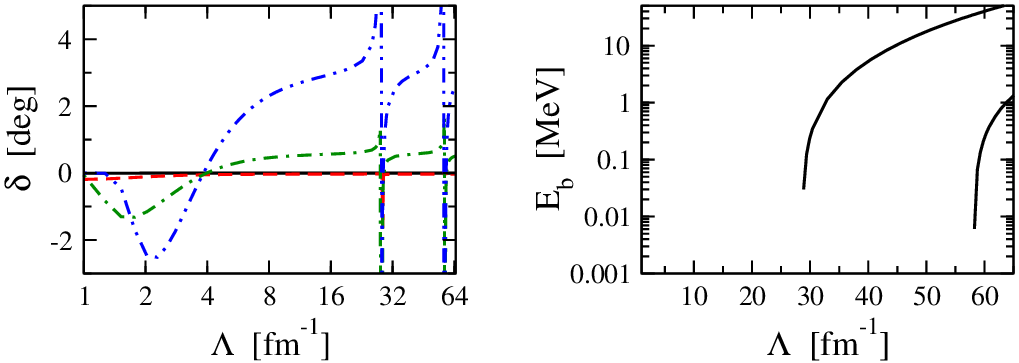}
    \caption{
         Cutoff dependence of the $^3D_3$ phase shift (left panel) and the
         binding energies of the spurious bound states (right panel). 
         Results for the phase shift are given for laboratory energies of 5 MeV (solid line), 
         25 MeV (short--dashed line), 100 MeV (dot--dashed line) and 200 MeV (dot--dot--dashed line). 
\label{3d3} 
 }
  \end{center}
\end{figure}

The
case of the $^3D_3$--$^3G_3$ channel requires a special consideration. 
Due to the relatively mild cutoff dependence for $\Lambda <
20$ fm$^{-1}$, no counterterm has been included in this channel in the
analysis of Ref.~\cite{Nogga:2005hy}. In
Fig.\ref{3d3} we show the dependence of the $^3D_3$ phase shift and of the
binding energies of the spurious bound states on the cutoff in the large range
for the values of the cutoff $\Lambda$.
The cutoff dependence of the phase shift is indeed rather small for $\Lambda <20$ 
fm$^{-1}$ at low energies but becomes significant at $E_{\rm lab} =$ 200
MeV. The first spurious bound state is generated in this channel at $\Lambda
\sim 29$ fm$^{-1}$. In order to study the behavior of the phase shifts in the 
limit $\Lambda \to \infty$ in these partial waves, we proceed in the same way 
as we did for the two other coupled channels and introduce a counterterm as
shown in the last line of Eq.~\ref{VC}. The corresponding LEC has to be
determined for a given value of $\Lambda$ from a fit to the $^3D_3$ phase
shift in the low--energy region. We found that the $^3D_3$ phase shift
starts to increase rapidly at energies of the order $E_{\rm
  lab} \sim 50$ MeV leading to dramatic deviations from the data if the fit is
performed in our standard range  $E_{\rm  lab} < 10$ MeV. This behavior can
probably be attributed to an unnaturally small value of this phase shift close
to threshold. In order to obtain meaningful results, we have, therefore,
decided to perform a global fit in the large energy region up to $E_{\rm  lab}
= 200$ MeV. 
\begin{figure}[tb]
  \begin{center} 
\includegraphics[width=16.6cm,keepaspectratio,angle=0,clip]{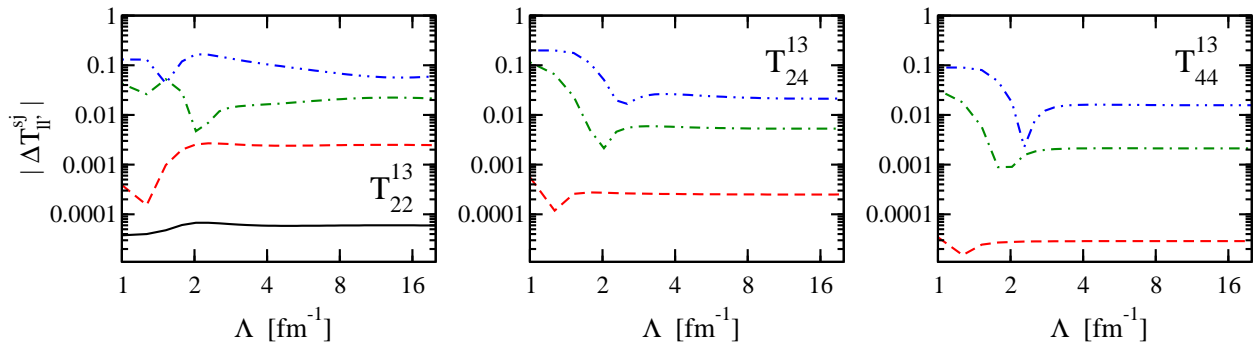}
    \caption{
         Error in T--matrix elements in the $^3D_3$--$^3G_3$ channel versus the cutoff $\Lambda$ 
         in the Lippmann--Schwinger equation. Results are given for laboratory energies of 5 MeV (solid line), 
         25 MeV (short--dashed line), 100 MeV (dot--dashed line) and 200 MeV (dot--dot--dashed line). 
\label{err_coup2} 
 }
  \end{center}
\end{figure}
Our results for $\Lambda$--dependence of the T--matrix elements obtained in
such a way are depicted in Fig.~\ref{err_coup2}. Similarly to the previously
discussed channels, keeping $\Lambda$ finite does not represent the main source
of uncertainty at LO already for $\Lambda \sim 2$ fm$^{-1}$. The only
exception is given by the $^3D_3$ phase shift at the highest energy
considered, $E_{\rm  lab}=200$ MeV, where the results at this order cannot be
trusted anyway. 

\begin{figure}[tb]
  \begin{center} 
\includegraphics[width=13.0cm,keepaspectratio,angle=0,clip]{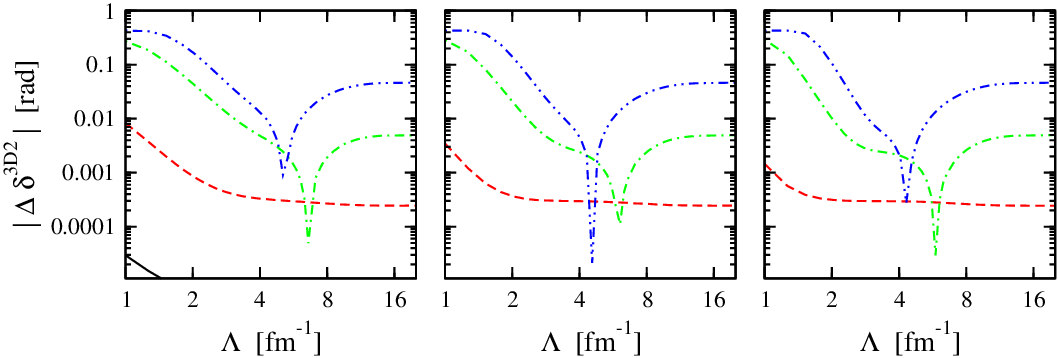}
    \caption{
         Error in the $^3D_2$ phase shift versus the cutoff $\Lambda$ 
         in the Lippmann--Schwinger equation. Left, middle and right pannels
         correspond to the choices $n=2$, 3 and 4 in Eq.~(\ref{reg_fun}).
         Results are given for laboratory energies of 5 MeV (solid line), 
         25 MeV (short--dashed line), 100 MeV (dot--dashed line) and 200 MeV (dot--dot--dashed line).
\label{err_3d2} 
 }
  \end{center}
\end{figure}

Finally, we have re-done the calculations using different regulator
functions. As a typical representative, we plot in Fig.~\ref{err_3d2} the
error in the $^3D_2$ phase shift 
versus the cutoff $\Lambda$ using $n=2$,  $n=3$ and  $n=4$ in the
regulator function in Eq.~(\ref{reg_fun}). Clearly, the critical value of the
cutoff is somewhat smaller for sharper regulator functions, but the conclusions
remain unchanged.  

\section{NN observables: Weinberg versus Nogga--Timmermans--van Kolck}
\def\theequation{\arabic{section}.\arabic{equation}}
\label{sec:observ}

Having convinced ourselves that a finite cutoff of the order $\Lambda \sim $ 2 -- 3
fm$^{-1}$ is sufficient in most partial waves to ensure that the cutoff dependence
is within the actual theoretical uncertainty of the LO
approximation,\footnote{Strictly speaking, in attractive channels this was
  only shown based on the power counting of Ref.~\cite{Nogga:2005hy}. We
  assume that this estimation is also valid for the Weinberg 
  counting. This assumption is supported by the results in the repulsive
  channels and by the qualitative arguments at the beginning of section III.}
we are
now in the position to study the implications for the power
counting. To be specific, we consider a selection of {\it np} scattering
observables at LO in chiral EFT with the finite cutoff chosen as described
above within the two different power counting schemes: the one of Weinberg (W)
and the one of Nogga et al. (NTvK). The potential in the first case is given by the OPEP
accompanied by two contact terms without derivatives that contribute to the
$^1S_0$ and $^3S_1$ partial waves. For the cutoff we use the value $\Lambda =
2.5$ fm$^{-1}$ which is close to the central value used in 
Refs.\cite{Epelbaum:2003xx,Epelbaum:2004fk} and has also been adopted in
\cite{Entem:2003cs}.  For the calculation based on the NTvK counting, we
choose $\Lambda = 3$ fm$^{-1}$ and include, in addition to the two S--wave
contact interactions also counterterms in the $^3P_0$, $^3P_2$ and $^3D_2$
channels as shown in Eq.~(\ref{VC}). For the sake of completeness, we will
also show the results in the NtvK counting scheme corresponding to $\Lambda =
20$ fm$^{-1}$, the largest cutoff considered in \cite{Nogga:2005hy}. 

\begin{figure}[tb]
  \begin{center} 
\includegraphics[width=16.6cm,keepaspectratio,angle=0,clip]{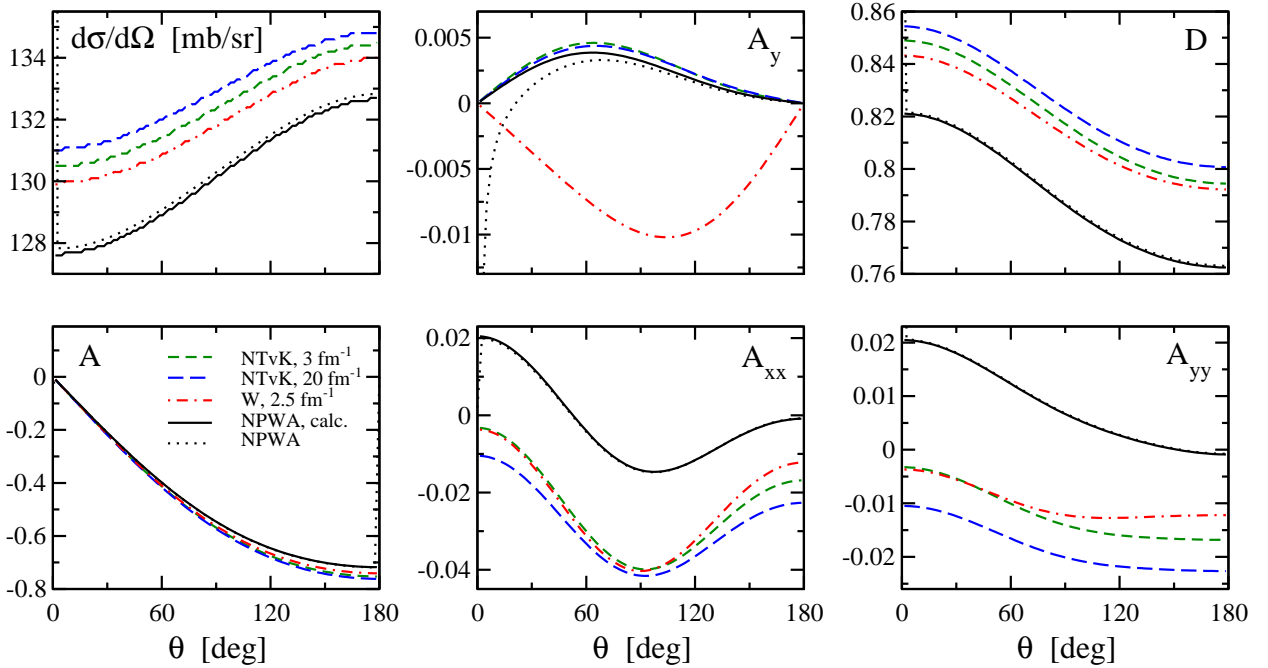}
    \caption{
         {\it np} differential cross section $d\sigma / d \Omega$, vector analyzing power $A_y$, 
         polarization transfer coefficients $D$ and $A$ and spin correlation parameters $A_{xx}$ and 
         $A_{yy}$ versus the scattering angle $\theta$ for laboratory energy of 5 MeV.  Short--dashed (long--dashed) 
         lines show the results obtained in the modified Weinberg power counting scheme with the cutoff $\Lambda = 3$ 
         fm$^{-1}$ ($\Lambda = 20$ fm$^{-1}$) while dot--dashed line depicts the results based on the original 
         Weinberg power counting. Solid lines shows the observables calculated based on phase shifts from 
         Nijmegen PWA while dotted line is the true Nijmegen PWA result from \protect{\cite{NNonline}}.   
\label{obs_5} 
 }
  \end{center}
\end{figure}

\begin{figure}[tb]
  \begin{center} 
\includegraphics[width=16.6cm,keepaspectratio,angle=0,clip]{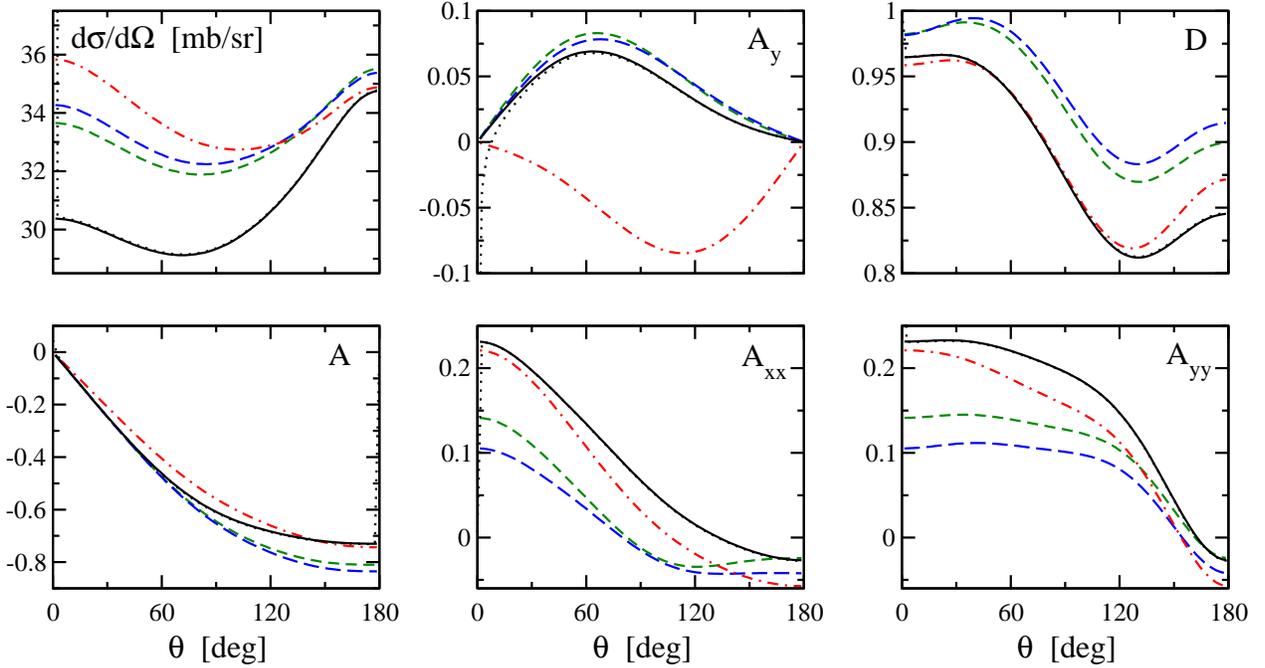}
    \caption{
         {\it np} differential cross section $d\sigma / d \Omega$, vector analyzing power $A_y$, 
         polarization transfer coefficients $D$ and $A$ and spin correlation parameters $A_{xx}$ and 
         $A_{yy}$ versus the scattering angle $\theta$ for laboratory energy of 25 MeV.  For remaining 
         notation see Fig.~\ref{obs_5}.
\label{obs_25} 
 }
  \end{center}
\end{figure}

\begin{figure}[tb]
  \begin{center} 
\includegraphics[width=16.6cm,keepaspectratio,angle=0,clip]{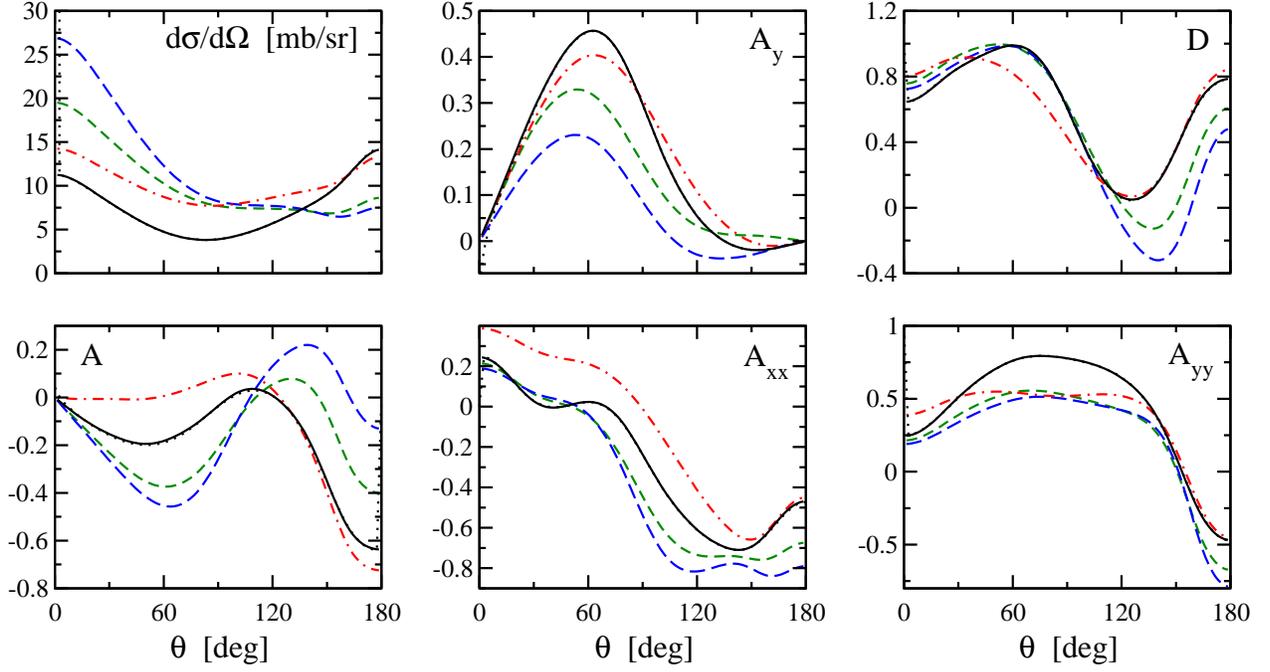}
    \caption{
         {\it np} differential cross section $d\sigma / d \Omega$, vector analyzing power $A_y$, 
         polarization transfer coefficients $D$ and $A$ and spin correlation parameters $A_{xx}$ and 
         $A_{yy}$ versus the scattering angle $\theta$ for laboratory energy of 100 MeV.  For remaining 
         notation see Fig.~\ref{obs_5}.
\label{obs_100} 
 }
  \end{center}
\end{figure}

\begin{figure}[tb]
  \begin{center} 
\includegraphics[width=16.6cm,keepaspectratio,angle=0,clip]{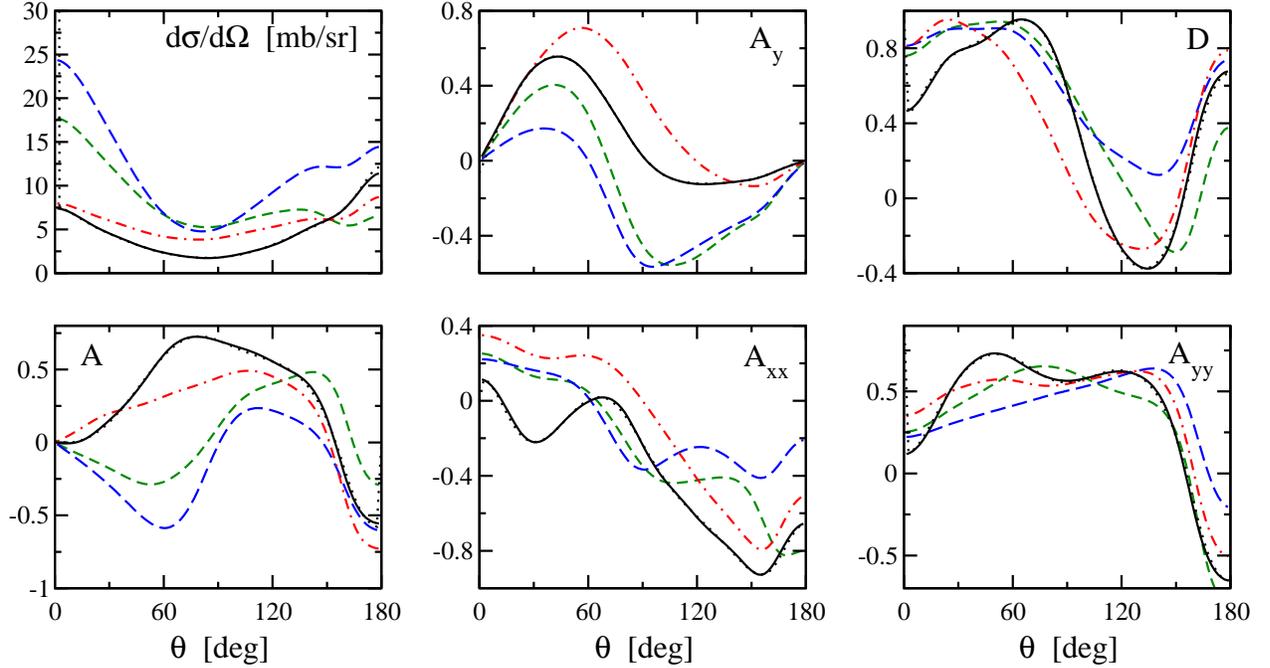}
    \caption{
         {\it np} differential cross section $d\sigma / d \Omega$, vector analyzing power $A_y$, 
         polarization transfer coefficients $D$ and $A$ and spin correlation parameters $A_{xx}$ and 
         $A_{yy}$ versus the scattering angle $\theta$ for laboratory energy of 200 MeV.  For remaining 
         notation see Fig.~\ref{obs_5}.
\label{obs_200} 
 }
  \end{center}
\end{figure}

\begin{figure}[tb]
  \begin{center} 
\includegraphics[width=11.0cm,keepaspectratio,angle=0,clip]{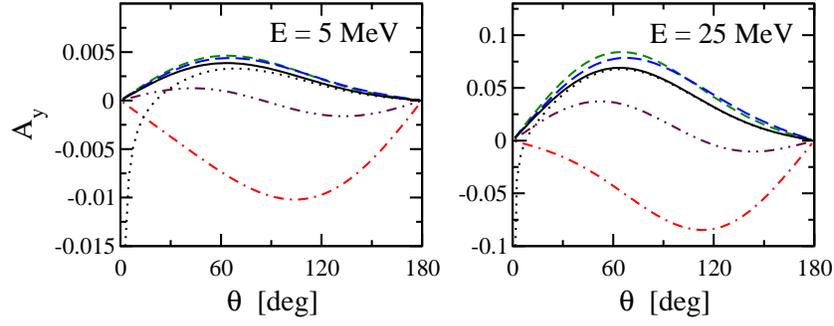}
    \caption{
         {\it np} vector analyzing power $A_y$ at $E_{\rm lab} = 5$ and 25~MeV.
         The double-dotted-dashed line shows the result in the W scheme with the
         $^3P_0$ partial wave taken from the Nijmegen PWA \protect{\cite{NNonline}}. For remaining 
         notation see Fig.~\ref{obs_5}.
\label{obs_ay} 
 }
  \end{center}
\end{figure}

The results for the differential cross
section, vector analyzing power $A_y$, depolarization observables $D$ and $A$
and spin correlation parameters $A_{xx}$ and $A_{yy}$ for
energies $E_{\rm lab} =$ 5, 25, 100 and 200 MeV are shown in Figs.\ref{obs_5}--\ref{obs_200}.
We use the same convention for observables as the one adopted by the Nijmegen
group \cite{NNonline}. For a precise definition of various nucleon--nucleon
scattering observables see e.g. the appendix of Ref.~\cite{Binstock:1974gv}.
The results obtained at LO in chiral EFT have to be compared to the ones obtained 
based on {\it np} phase shifts from Nijmegen PWA \cite{NNonline} which should be
regarded as experimental data and are shown by the solid lines. In the present
study, we neglect the electromagnetic interaction between the neutron and
proton. To visualize the effects due to the (neglected) magnetic moment
interaction, we also show by the dotted line the results for various
observables calculated by the Nijmegen group \cite{NNonline} which include the magnetic moment
interaction.  The only significant effects are observed for
$A_y$ at small scattering angles and energies as well as for differential
cross section in the forward direction.   
At the smallest energy, $E_{\rm lab}=5\,$MeV, all three EFT calculations lead
to similar results for $ d \sigma /d \Omega$, $D$ and $A$. In the case of vector
analyzing power $A_y$, the results based on the NTvK counting are consistent with
the data while the ones obtained in the W scheme do not even reproduce the
correct sign. This failure
of the W counting can be traced back to the relatively large error in the
$^3P_0$ partial wave at low energy, see Fig.\ref{obs_ay}, and to the
well--known extreme sensitivity of $A_y$ to the triplet P--waves. In the NTvK
counting, the $^3P_0$ is, by construction, much better reproduced at low energy due to the
presence of the additional counter term. We emphasize, however, that despite
the large relative error, the absolute error for $A_y$ in the W scheme is
still small as a consequence of its tiny size. Similarly, large relative deviations of the
order of 100\% are observed in both W and NTvK counting schemes for $A_{xx}$
and  $A_{yy}$, which are also rather small at this energy.  

The results at $E_{\rm lab} = 25$ MeV are shown in Fig.~\ref{obs_25}. The
differential cross section is better reproduce in the NTvK scheme in the forward
and in the W scheme in the backward angles. The situation with the vector
analyzing power, which is still very small numerically, is similar to the one
at  $E_{\rm lab} = 5$ MeV. On the other hand the spin correlation parameters
$A_{xx}$ and $A_{yy}$, which have a more natural size, are now much better
reproduced.  Notice that for $D$, $A$, $A_{xx}$ and  $A_{yy}$, the results
obtained in the W counting show a significantly better agreement with the data
compared to the  ones based on the NTvK scheme.  We also emphasize that
increasing the cutoff in the NTvK scheme from $3$ to $20$ fm$^{-1}$ does not
lead to improvement, which is consistent with our
findings in section  \ref{sec:phases}.

At $E_{\rm lab} = 100$ MeV, one observes moderate (large) deviations for the differential
cross section in the W (NTvK) counting, see Fig.~\ref{obs_100}.  
The description of the data in the
NTvK scheme gets worse with increasing $\Lambda$. The situation with the
analyzing power is now opposite to the one observed at the two smaller
energies. It is best described in the W scheme, while the prediction in the
NtvK scheme deviates strongly from the data, especially in the case of the
large cutoff. The results for $D$, $A_{xx}$ and $A_{yy}$ are comparable in both
approaches while the description of $A$ is slightly better in the W counting. 

Finally, at $E_{\rm lab} = 200$ MeV, the largest energy considered, the
results start to diverge even for the differential cross
section, see Fig~\ref{obs_200}. Clearly, no quantitative agreement with the
data can be expected at this high energy at LO in the chiral expansion.
The prediction for $d\sigma / d \Omega$ based on the W counting turns out to be
in a much better agreement with the data compared to the NTvK scheme. The
results for all other observables are of a similar quality and show mostly a 
qualitative agreement with the data. 

\begin{figure}[tb]
  \begin{center} 
\includegraphics[width=16.6cm,keepaspectratio,angle=0,clip]{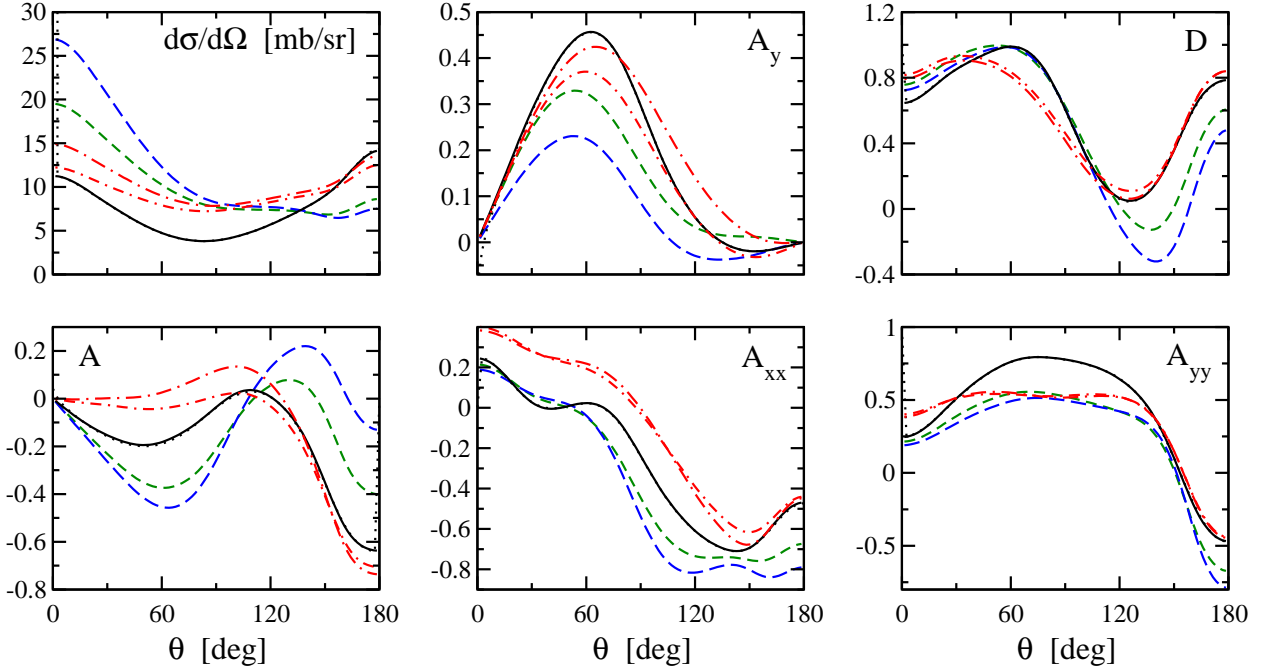}
    \caption{
         {\it np} differential cross section $d\sigma / d \Omega$, vector analyzing power $A_y$, 
         polarization transfer coefficients $D$ and $A$ and spin correlation parameters $A_{xx}$ and 
         $A_{yy}$ versus the scattering angle $\theta$ for laboratory energy
         of 100 MeV.  The short- (long-) dashed lines correspond to choosing $n=2$
         ($n=4$) in Eq.~(\ref{reg_fun}). For remaining 
         notation see Fig.~\ref{obs_5}.
\label{obs_100_reg} 
 }
  \end{center}
\end{figure}

We have also carried out the calculations in the W scheme using the regulator functions 
in Eq.~(\ref{reg_fun}) with $n=2$ and $n=4$. The results for observables at
$E_{\rm lab} = 200$ MeV are depicted in Fig.~\ref{obs_100_reg} together with
the results in the NTvK approach. 

Based on the above results, we conclude that both W and NTvK 
counting schemes lead to a similar description of the {\it np} scattering data
for the appropriately chosen cutoff (i.e.~$\Lambda \sim 3$ fm$^{-1}$).  
The modification of the original W counting by promoting certain
higher--order contact interactions to LO in the NTvK scheme seems not to be 
supported by the data. In addition, we confirm our findings in the 
previous section that increasing $\Lambda$ in the calculations based 
on the NTvK scheme does not yield any improvement.

\section{Generalization to higher orders}
\def\theequation{\arabic{section}.\arabic{equation}}
\label{sec:discussion}

\begin{figure}[tb]
  \begin{center} 
\includegraphics[width=16.0cm,keepaspectratio,angle=0,clip]{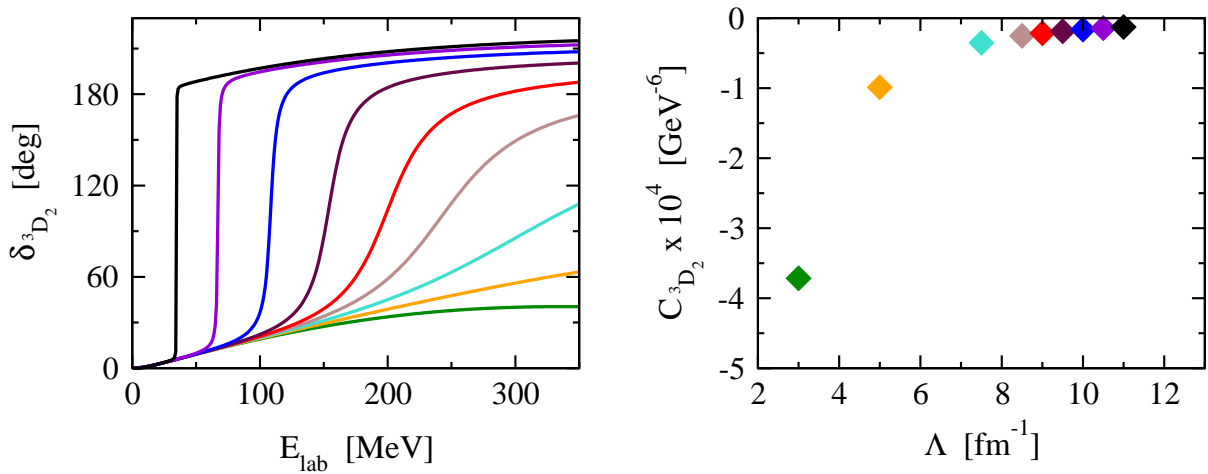}
    \caption{
         $^3D_2$ phase shift obtained by solving the LS equation with the one--,
         the leading two--pion exchange potential and the counterterm from
         Eq.~(\ref{VC}) (left panel). The lines (from bottom to top)
         correspond to $\Lambda=$ 3.0,  5.5, 7.5, 8.5, 9.0, 9.5, 10.0, 10.5
         and 11.0 fm$^{-1}$. In the right the cutoff dependence of the
         corresponding LEC versus the cutoff $\Lambda$ is shown.
\label{3d2_nlo} 
 }
  \end{center}
\end{figure}

Let us now briefly discuss some possible ways to include higher-order corrections in
the NTvK scheme, see Ref.~\cite{Valderrama:2005wv} for a much more detailed discussion. 
A straightforward generalization by solving the LS equation
with the  potential which includes higher-order chiral corrections 
appears, in
general, not to possible. 
To see that consider e.g.~the $^3D_2$ partial wave
at NLO. Since the OPEP is singular and attractive in this channel, the
LO potential in the NTvK scheme includes, in addition, one counterterm as shown in
Eq.~(\ref{VC}). At NLO, the long-range part of the potential receives the
correction due to the leading $2\pi$--exchange, which is repulsive in this
channel and behaves at short distances as $1/r^5$. We now follow the 
approach of section \ref{sec:phases} and determine the value of $C_{^3D_2}$ as a
function of $\Lambda$ from a fit to the Nijmehen phase shift for $E_{\rm lab}
< 10$ MeV. As shown in Fig.\ref{3d2_nlo}, the nonperturbative inclusion of
the leading $2\pi$--exchange potential leads to the appearance of a resonance,
which gets sharper and moves to lower energies with increasing the cutoff
$\Lambda$. Notice that we did not observe bound states in this
channel even for $\Lambda$'s significantly larger than the ones shown in
Fig.\ref{3d2_nlo}. 
This example clearly indicates that such an approach  becomes untunable
for cutoff values $\Lambda \gtrsim 10$ fm$^{-1}$. In this context, we would
also like to emphasize that certain channels were found to be problematic at
NLO in the approach by Pavon Valderrama and Ruiz Arriola based on $\Lambda \to
\infty$ \cite{Valderrama:2005wv,PavonValderrama:2005uj}.\footnote{Notice that
while the above mentioned difficulties are absent at N$^2$LO due to the strong
attraction caused by the isoscalar central part of the subleading TPEP
\cite{Valderrama:2005wv,PavonValderrama:2005uj}, the
appearance of the singular repulsive potentials is unavoidable at higher
orders in the chiral expansion. } 

The above mentioned difficulties clearly demonstrate that a generalization of
the NTvK scheme to include higher--order chiral corrections is not
straightforward. Possible scenarios include:
\begin{itemize}
\item[i)]
One may try to promote additional higher--order counterterms to in order to remove
resonances from the low--energy region. This scenario seems to be in
conflict with the $\Lambda \to \infty$ formulation based on the boundary
condition approach of \cite{Valderrama:2005wv,PavonValderrama:2005uj}. 
\item[ii)]
The approach by Pavon Valderrama and Ruiz Arriola 
\cite{Valderrama:2005wv,PavonValderrama:2005uj} 
allows to avoid the above--mentioned problems by demoting
the counter terms in the singular repulsive channels. Such an approach, however, 
appears to be not systematically improvable in the sense that it
does not provide a systematic way to include short--range physics beyond the
pion-exchange contributions. For a recent study along these lines supporting the
conclusions of this work see \cite{Zeoli:2012bi}. 
\item[iii)]
Finally, there is a more natural solution based upon a perturbative treatment
of the higher--order corrections, which reduces to the distorted wave
perturbation theory using the exact LO wave functions
\cite{Nogga:2005hy,Birse:2005um}. As pointed out in \cite{Valderrama:2005wv}, this would
require a large number of counterterms, which raises the concern about its
predictive power,  for the details. It remains to
be seen whether such an scheme will lead to a perturbative expansion for the
nucleon--nucleon observables.
\end{itemize}

\section{Summary and conclusions}
\def\theequation{\arabic{section}.\arabic{equation}}
\label{sec:summary}

We summarize by repeating  the logic we have followed in this manuscript:

\begin{itemize}
\item[i)]
The LS (or Schr\"odinger) equation with  the point--like static OPEP is
ill-defined (ultraviolet divergent). An infinite number of counterterms are needed 
in the limit $\Lambda \to \infty$. Treating the high partial waves in perturbation
theory cannot be justified if one does not make assumptions about the
range of applicability of the point-like OPEP, i.e. if one does not implicitly
introduces a finite UV cutoff. The approach with $\Lambda \to \infty$ seems
therefore to be useless if one wants to consider observables rather than
specific partial waves. 
\item[ii)]
Of course, this is not a conceptual problem in an EFT. The potential
description is only valid for small momenta/energies. There are infinitely many 
forms of the OPEP potential which differ from each other by short-range
terms and that are equivalent from the viewpoint of EFT. Choosing an appropriate
short-range extension corresponds to using a finite cutoff. A striking example
is the relativistically corrected expression for OPEP, cf. Eq.~(\ref{opep_rel})
-- it does not lead to divergences in the LS equation.  For an 
extended discussion on nonperturbative renormalization and the role of
the cutoff the reader is referred to
Refs.~\cite{Epelbaum:2009sd,Epelbaum:2012ua} and references therein.    
\item[iii)]
If one agrees to work with a finite $\Lambda$, than the question is how to
choose it. Our answer is: \emph{Any value of $\Lambda$ is acceptable if the
error associated to its finite value is within the theoretical uncertainty 
at a given order.} This is a well--defined criterion in contrast to the one 
used in Ref.~\cite{Nogga:2005hy} (called ``strong'' or ``mild'' cutoff
dependence). We have studied the NN phase shifts based on the OPEP
and found that starting from $\Lambda \sim 3$ fm$^{-1}$, the error due to
keeping $\Lambda$ finite is within the theoretical uncertainty at LO. 
\item[iv)]
It is advantageous for many reasons to keep $\Lambda$ small (yet acceptable),
see \cite{Lepage:1997,Lepage:2000,Gegelia:2004pz,Bogner:2006tw}.   
\item[v)]
Having convinced ourselves that $\Lambda \sim 3$ fm$^{-1}$ is acceptable, the
question is: what counterterms need to be included? The choices are: i) 
W--counting (based on  naive dimensional analysis); ii) NTvK counting (based
on the asymptotics in the limit $\Lambda \to \infty$). We have compared the
predictions for various observables in these two counting schemes with the finite
cutoff choosen as discussed above and found no improvements due to 
the promotion of certain counterterms to LO as suggested in the NTvK
scheme (except for the case of $A_y$ as discussed above). Increasing the
cutoff to $\Lambda = 20$ fm$^{-1}$ in the NTvK scheme makes the description of
the data in most cases even worse.  
\end{itemize}

Finally, we list a few advantages of the W--approach with a suitably chosen
finite cutoff $\Lambda$:

\begin{itemize}
\item[i)]  
The potentials constructed so far are fairly smooth for the ranges of
cutoffs considered, this implies that few-- and many--body calculations
are simpler to perform since one does not have to deal explicitely with spurious
deeply bound states. In particular, this leads to ``more perturbative''
many-body calculations, see e.g. the discussion in Ref.~\cite{Bogner:2006tw}.
\item[ii)] At a given order, one has less parameters, that means that the  
predictive power is higher. Furthermore, we stress again that the choice of 
the cutoff assumes the knowledge of the separation scale.
\item[iii)] The Weinberg approach can directly be extended to systems for $N$ 
($N > 2$) nucleons based on the  $N$--nucleon  Schr\"odinger equation with no
need to deal with spurious deeply bound states. 
These forces are also amenable to a lattice formulation, as recently shown in
Ref.~\cite{Epelbaum:2010xt,Epelbaum:2011md}.

\item[iv)] Finally, one might pose the question whether the mild explicit
cutoff dependence observed in higher order calculations in the Weinberg
scheme really is to be considered a problem?
\end{itemize}

We stress again that to really make progress on these issues, it is mandatory
to go beyond leading order - nuclear physics can not be described by one--pion
exchange plus a certain number of contact terms. In particular, we challenge
the authors of Ref.~\cite{Nogga:2005hy} to present a  viable and
phenomenological successful scheme that could truly be considered an 
alternative to Weinberg's original proposal.

\section*{Acknowledgments}

We would like to thank Andreas Nogga for helpful comments on the manuscript. 
This work has been supported by the Helmholtz Association, contract number VH-NG-417.
Partial financial support from  the EU Integrated Infrastructure
Initiative HadronPhysics3 Project, the European Research Council (ERC-2010-StG 259218 NuclearEFT), the DFG and the NSFC through
funds provided to the Sino-German CRC 110 ``Symmetries and
the Emergence of Structure in QCD'' and the BMBF (06BN7008)
is gratefully acknowledged.

\setlength{\bibsep}{0.2em}
\bibliographystyle{h-physrev3.bst}
\bibliography{/Users/ee/ee_vaio_linux/jlab/refs_h-elsevier3}

\end{document}